\documentclass[12pt]{article}
\usepackage{amssymb,amsmath,amsfonts}

\def\ba{\begin{array}}
\def\ea{\end{array}}

\def\ben{\begin{enumerate}}
\def\een{\end{enumerate}}
\def\beqan{\begin{eqnarray*}}
\def\eeqan{\end{eqnarray*}}
\def\btab{\begin{tabular}}
\def\etab{\end{tabular}}
\def\bit{\begin{itemize}}
\def\eit{\end{itemize}}

\newcommand{\lra}{\longrightarrow}

\newcommand{\mbb}[1]{\mathbb{#1}}
\newcommand{\C}{{\mbb C}}
\newcommand{\CP}{{\C\mbb P}}
\newcommand{\Z}{{\mbb Z}}
\newcommand{\R}{{\mbb R}}

\begin{document}

\topmargin -2pt

\headheight 0pt

\topskip 0mm \addtolength{\baselineskip}{0.20\baselineskip}
\begin{flushright}
{\tt hep-th/0204013}
\end{flushright}

\vspace{10mm}

\begin{center}
{\large \bf   N-Point Deformation of Algebraic K3  Surfaces}\\

\vspace{7mm}

Hoil Kim\footnote{hikim@gauss.knu.ac.kr}\\

{\it Topology and Geometry Research Center, Kyungpook National University,\\
Taegu 702-701, Korea}\\

\vspace{3mm}

and \\

\vspace{3mm}

Chang-Yeong Lee\footnote{cylee@sejong.ac.kr}\\

{\it Department of Physics, Sejong University, Seoul 143-747, Korea}\\

\vspace{18mm}

\end{center}

\begin{center}
{\bf ABSTRACT}
\end{center}
We consider 
N-point deformation of algebraic K3 surfaces.
First, we construct two-point deformation of algebraic K3 surfaces 
by considering algebraic deformation of 
a pair of commutative algebraic K3 surfaces.
In this case, the moduli space of
the noncommutative deformations is of dimension 19, 
the same as the moduli dimension of the complex 
deformations of commutative algebraic K3 surfaces.
Then, we extend this method to the N-point case.
In the N-point case, the dimension of deformation moduli space 
becomes 19N(N-1)/2.  
\\

\vfill


\thispagestyle{empty}

\newpage
\section*{I. Introduction}

Ever since the work of Connes, Douglas, and Schwarz \cite{cds} 
connecting the noncommutative torus and the T-duality in the M theory context
appeared in the string/M theory arena, the field related with
noncommutaive geometry \cite{conn} becomes an industry in the string/M theory circle.
Notably, noncommutative torus \cite{cr,rief} and its varieties have been studied intensively
\cite{nct,hv,t4ours,sw}. However, noncommutative versions of the K3 surfaces and the
 Calabi-Yau(CY) threefolds have been rarely studied \cite{bl1,kl1,bs} (see also \cite{ks1,ks2,kkl}).
Only recently, noncommutative tori with complex structures have been studied \cite{schwarz,manin}.

In Ref.\cite{bjl}, Berenstein, Jejjala, and Leigh initiated an algebraic geometry approach
to noncommutative moduli space. Then in Ref.\cite{bl1}, 
Berenstein and Leigh discussed noncommutative CY threefold from the
viewpoint of algebraic geometry. They considered two examples: a toroidal orbifold $T^6/{\Z}_2 \times
{\Z}_2$ and an orbifold of the quintic in ${\CP}^4$, each with
discrete torsion \cite{vafa,vfwt,doug,dgfl,jgomi}.
There, they explained the fractionation of branes at singularities
 from noncommutative geometric viewpoint under the presence of
 discrete torsion.

In Ref.\cite{bl1},  Berenstein and Leigh considered the $T^6/{\Z}_2 \times {\Z}_2$ case
 and recovered a large slice of the moduli space of complex structures of
 the CY threefold
 from the deformation of the noncommutative resolution of the orbifolds
via central extension of the local algebra of holomorphic
 functions.
In the commutative K3 case, the moduli space for the K3 space itself has been known already
 (see for instance \cite{aspinwall}),
and even the moduli space for the bundles on K3 surfaces
 has been studied \cite{mukai}.
In the noncommutative deformation of CY threefolds in Ref.\cite{bl1},  
the three holomorphic coordinates $y_i$
anticommute with each other to be compatible with ${\Z}_2$ discrete torsion.

In our previous work \cite{kl1}, we applied this algebraic approach to K3 surfaces 
 in the cases of the orbifolds $T^4/{\Z}_2$. 
We constructed a family of
noncommutative K3 surfaces by algebraically deforming  $T^4/{\Z}_2$
in both complex and noncommutative directions altogether.
In that construction the dimensions of moduli spaces
for the complex structures and the noncommutative deformations
were the same 18, which is the dimension of the moduli space
of the complex structures of K3 surfaces constructed with two elliptic
curves.

However, in the commutative case
the complete family of  complex deformations of K3 surfaces is of 20 dimension
inside which that of the algebraic K3 surfaces is of 19 dimension \cite{aspinwall}.
Thus, in this papar, we first construct a 19 dimensional
family of the noncommutative moduli of  
general algebraic K3 surfaces by considering algebraic deformation of
a pair of K3 surfaces. 
This construction apparently looks similar to 
the Connes-Lott's ``two-point space" construction of the standard model \cite{cl90}.
Thus, we will call it ``two-point deformation".
Next, we extend this method directly to the N-point case
by deforming N-tuple of commutative algebraic K3 surfaces
embedded in 
${\mbb P}^2 (x_1, x_2) \times {\mbb P}^1 (t_1) \times \cdots \times {\mbb P}^1 (t_N)$.

In section II, we construct a two-point deformation for
general algebraic K3 surfaces.
In section III, we 
extend the method to the N-point case.
In section IV, we conclude with discussion.

\section*{II. Two-point deformation }\label{two-point}

In this section, we first
 consider the ``two-point space" version of noncommutative deformation for 
general algebraic K3 surfaces in the direct extension of our previous work on
noncommutative $T^4/{\Z}_2$ \cite{kl1}.
General algebraic K3 surfaces are given by the following form
and with a point added at infinity.
\begin{equation}
\label{algk3}
y^2 = f(x_1, x_2)
\end{equation}
Here $f$ is a function with total degree 6 in $x_1,x_2$.

Now, we compare this with the Kummer surface,
the orbifold of $T^4/{\Z}_2$ case \cite{kl1}. 
 There we considered $T^4$ as the product of two elliptic curves, each given in Weierstrass form
\begin{equation}
\label{t2}
y_i^2 = x_i (x_i-1)(x_i - a_i)
\end{equation}
with a point added at infinity for $i=1,2.$
By the following change of variables, the point at infinity is brought to a finite point:
\begin{eqnarray}
\label{t2tr}
y_i \lra y_i' = \frac{y_i}{x_i^2}, \\
x_i \lra x_i' = \frac{1}{x_i} . \nonumber
\end{eqnarray}


For algebraic K3 surfaces, we first consider a function with total degree 6
in complex variables $u,v,w,$ for instance
\[ F(u,v,w) =u^2v^3w + u^4v^2 . \]
In a patch where the point at infinity of $w$ can be brought to a finite point, 
this can be written as
\[ \frac{F}{w^6} = (\frac{u}{w})^2 (\frac{v}{w})^3 + (\frac{u}{w})^4 (\frac{v}{w})^2 \]
and may be denoted as
\[ f(x_1,x_2) =x_1^2 x_2^3 +x_1^4 x_2^2  \]
where $x_1=\frac{u}{w}, x_2=\frac{v}{w} .$
Then, an algebraic K3 surface is given by
\begin{equation}
\label{algk3p1}
y^2 = f(x_1, x_2) = x_1^2 x_2^3 +x_1^4 x_2^2  .
\end{equation}
Similarly, in a patch where the point at infinity of $u$ can be brought to a finite point,
we consider a function
\[ \frac{F}{u^6} = (\frac{v}{u})^3 \frac{w}{u} + (\frac{v}{u})^2 , \]
and this can be written as
\begin{equation}
\label{algk3p2}
{y'}^2 = f'(x_1', x_2') = {x_1'}^3 x_2' + {x_1'}^2 
\end{equation}
where $x_1'=\frac{v}{u} = \frac{x_2}{x_1}, x_2' =\frac{w}{u} = \frac{1}{x_1} . $
This can be also obtained directly from (\ref{algk3p1}) by dividing it with $x_1^6$
\[ \frac{y^2}{x_1^6} =  \frac{x_1^2 x_2^3}{x_1^6} + \frac{x_1^4 x_2^2}{x_1^6}
         =   (\frac{x_2}{x_1})^3 \frac{1}{x_1} + (\frac{x_2}{x_1})^2   . \] 
Thus, in the case of the general algebraic K3, a point at infinity in one patch 
can be brought to a finite point in another patch by the following change of
variables 
\begin{eqnarray}
\label{k3ytr}
y \lra y' = \frac{y}{x_1^3}, \\
\label{k3xtr}
x_1 \lra x_1' = \frac{x_2}{x_1} , \\ 
x_2 \lra x_2' = \frac{1}{x_1} .
\nonumber
\end{eqnarray}

We now consider a deformation of algebraic K3 surfaces in noncommutative 
direction.
Following the same reasoning in our previous work \cite{kl1},
we consider two commuting complex variables $x_1, x_2$ and 
two noncommuting variables $t_1, t_2$ such that
\begin{eqnarray}
\label{k3ftn}
t_1^2 = h_1(x_1, x_2), \\
t_2^2 = h_2(x_1, x_2),
\nonumber
\end{eqnarray}
where $h_1, h_2$ are commuting functions of total degree 6 in $x_1, x_2$.
To be consistent with the condition that
 $t_1^2, t_2^2$ belong to the center, one can allow
the following deformation for $t_1, t_2$.
\begin{equation}
\label{Pctr}
t_1 t_2 + t_2 t_1 = P(x_1, x_2)
\end{equation}
Here the right hand side should be a polynomial and free of poles in each patch.
Thus, under the change of variables (\ref{k3xtr})
\begin{eqnarray*}
x_1 \lra x_1' = \frac{x_2}{x_1} , \\ 
x_2 \lra x_2' = \frac{1}{x_1}, 
\end{eqnarray*}
$t_i$ should be changed into
\begin{equation}
\label{ttr}
t_i  \lra t_i' = \frac{t_i}{x_1^3},  \  \  \  \  {\rm for} \ \  i=1,2  . 
\end{equation}
This is due to the fact that $t$'s transform just like $y$ in (\ref{k3ytr}).
Therefore, $P$ transforms as
\begin{equation}
\label{Ptr}
P(x_1, x_2)  \lra  x_1^6 P' ( \frac{x_2}{x_1},  \frac{1}{x_1} ).  
\end{equation}
This implies that $P'$ should be of total degree 6 in $x_1', x_2'$, at most. 
Interchanging the role of $P$ and $P'$ one can see that $P$ should  be also
of total degree 6 in $x_1, x_2$. 

The above structure can be understood in the following manner.
If we do not impose the condition (\ref{Pctr}), and if we have only one of $t_i$'s
satisfying the condition (\ref{k3ftn}), then we have only one copy of an algebraic
K3 surface.
If we have both $t_i$'s without the condition (\ref{Pctr}), then we have two copies
of K3 surfaces. If we have both $t_i$'s and impose the condition (\ref{Pctr}),
then we have a noncommutatively deformed K3 surfaces 
in which the above mentioned two K3 surfaces intertwined each other  
everywhere on their surfaces, becoming fuzzy. 
This seems to be similar to the ``two-point space" version of the Connes-Lott model \cite{cl90}.
In the Connes-Lott model, every point of the space
becomes fuzzy due to the 1-to-2 correspondence at each point in the space,
where the two corresponding points at each classical location are pre-fixed.
On the other hand, ours are more or less like position $x$ and momentum $p$ 
in quantum mechanics at every point in the space.
However, since we started with two copies of the classical space just like the Connes-Lott
model, and combined them
to become a noncommutative space, we will call our construction  
 ``two-point deformation" 
though our construction is not exactly the same as the Connes-Lott's in its nature.

Now, we count the dimension of the moduli space of our deformation.
In our previous work for noncommutative $T^4/{\Z}_2$ \cite{kl1},
 $t_1$ for $y_1y_2$ and $t_2$ for $y_2y_1$ were all invariants of the K3
surface.
The dimensions of the moduli spaces of these deformations were 18
for both the noncommutative and complex deformation cases, matching
the moduli space dimension of the complex deformation for  $T^4/{\Z}_2$ .
In the present case, from eq.(\ref{Ptr}) we can see
that the dimension of the moduli spaces of these deformations are
19 for both the noncommutative and complex deformation cases. 
In fact, to show this we need to count the dimension of the
polynomials of degree 6 in three variables up to constant modulo
projective linear transformations of three variables. We get
$19=28-1-8$, where 28 is the dimension of polynomials of degree 6 in
three variables and 1 and 8 correspond to a constant and
$PGL(3, \C)$, respectively.

\section*{III. N-point deformation }\label{cp3}

In this section, we follow the method in the previous section and
 consider the ``N-point space" of the noncommutative deformation of the 
general algebraic K3 surfaces.

  First, we consider $N$-tuple of commutative algebraic K3 surfaces
\begin{eqnarray}
\label{nk3ftn}
t_1^2  & = & h_1(x_1, x_2), \nonumber \\ 
 & \vdots &    \\ 
t_N^2 & =  & h_N(x_1, x_2),
\nonumber
\end{eqnarray}
where $h_1, \cdots,  h_N$ are commuting functions of total degree 6 in $x_1, x_2$.
This can be regarded as embedding the i-th copy of algebraic K3 surface $X_i$ in
${\mbb P}^2 (x_1, x_2) \times {\mbb P}^1 (t_i)$ as 
$t_i^2=h_i(x_1, x_2)$ of a degree 6 polynomial.
Locally the algebra representing the functions on $X_i$ can be expressed as
${\C} [x_1, x_2, t_i]/I_i$, where $I_i$ is a principal ideal generated by
$t_i^2 -h_i(x_1, x_2)$ and 
 ${\C} [x_1, x_2, t_1, \cdots, t_N]$ is a local polynomial algebra of 
${\mbb P}^2 (x_1, x_2) \times {\mbb P}^1 (t_1) \times \cdots \times {\mbb P}^1 (t_N)$.
Thus embedding   $X_i$ in
${\mbb P}^2 (x_1, x_2) \times {\mbb P}^1 (t_1) \times \cdots \times {\mbb P}^1 (t_N)$
induces a natural quotient map from
${\C} [x_1, x_2, t_1, \cdots, t_N]$  to ${\C} [x_1, x_2, t_i]/I_i$ by putting
$t_j$ as 0  for $j \neq i$.

Now, we consider the deformation of this embedded space in the noncommutative direction
as in the two-point case. 
In order to be consistent with the condition that
 $t_1^2, \cdots, t_N^2$ belong to the center along with $x_1, x_2$, 
we can allow the following deformation for $t_1, \cdots,  t_N$.
\begin{equation}
\label{nPctr}
t_i t_j + t_j t_i = P_{ij}(x_1, x_2), \  \  \  \  {\rm for} \ \  i,j=1, \cdots, N, \ \ i \neq j. 
\end{equation}
Here the right hand side should be a polynomial and free of poles in each patch.
Thus, under the change of variables (\ref{k3xtr})
\begin{eqnarray*}
x_1 \lra x_1' = \frac{x_2}{x_1} , \\ 
x_2 \lra x_2' = \frac{1}{x_1}, 
\end{eqnarray*}
$t_i$'s should be changed into
\begin{equation}
\label{nttr}
t_i  \lra t_i' = \frac{t_i}{x_1^3},  \  \  \  \  {\rm for} \ \  i=1, \cdots, N. 
\end{equation}
This is due to the fact that $t_i$'s transform just like $y$ in (\ref{k3ytr}).
Therefore, $P_{ij}$ transforms as
\begin{equation}
\label{nPtr}
P_{ij}(x_1, x_2)  \lra  x_1^6 P_{ij}' ( \frac{x_2}{x_1},  \frac{1}{x_1} ).  
\end{equation}
This implies that $P_{ij}'$ should be of total degree 6 in $x_1', x_2'$, at most. 
Interchanging the role of $P_{ij}$ and $P_{ij}'$ one can see that $P_{ij}$ should  be also
of total degree 6 in $x_1, x_2$. 

If we forget the embedded $N$ K3 surfaces given by the constraints (\ref{nk3ftn}) 
for the time being,
the above defined $ \{ P_{ij}(x_1, x_2) \} $ 
given by (\ref{nPctr}) define a deformation of the ambient space 
  ${\mbb P}^2 (x_1, x_2) \times {\mbb P}^1 (t_1) \times \cdots \times {\mbb P}^1 (t_N)$.
So, we can understand that imposing the condition of the change of chart (\ref{nttr}),(\ref{nPtr})
compatible to the complex structures coming from (\ref{nk3ftn})  induces a
restriction on $P_{ij}$ being of total degree 6 in $x_1, x_2$.
We might call this deformation a deformation of $N$  K3 surfaces.
The choice of $P_{ij}$ is independent of the choice of $h_i$, which 
means that the deformations of the classical complex structure and
of the noncommutative structure are independent of each other as expected.
Now, we count
the dimension of the moduli space of our deformation.
In the two-point case of the previous section, 
the dimension of the moduli space of the deformation was
19.
In that case, 
we counted the dimension of the
polynomials of degree 6 in three variables up to constant modulo
projective linear transformations of three variables. Thus, we got
$19=28-1-8$, where 28 is the dimension of polynomials of degree 6 in
three variables and 1 and 8 correspond to a constant and
$PGL(3, \C)$, respectively.
Thus, in the N-point case the dimension of the moduli space
of the deformation is the number of independent $P_{ij}$ times
the deformation dimension of the two-point case.
Namely, we have 19N(N-1)/2 as the dimension of the deformation moduli
for the N-point deformation case.



\section*{IV. Discussion}

In this paper,
we deformed 
$N$ K3 surfaces in the noncommutative sense and
computed the dimension for the moduli space. 

In the first part of the paper, 
we constructed  the
two-point deformation of algebraic K3 surfaces 
by considering algebraic deformation of 
a pair of commutative algebraic K3 surfaces.
Doing this, 
we used the same method as in the case of 
the Kummer K3 surface\cite{kl1} which is the ${\Z}_2$
quotient of two elliptic curves $E_1, E_2$ where
$E_i$ satisfies $ y_i^2 =f_i (x_i)$.
In Ref.\cite{kl1}, we defined 
$t_1=y_1y_2,$  $t_2 =y_2y_1$ and introduced
the deformation 
$t_1t_2 +t_2 t_1= P_{12}(x_1, x_2)$.
In that case $t_1, \ t_2$ were functions on the Kummer K3 surface, so 
that the deformation was a noncommutative deformation
of one Kummer K3 surface. However, 
the moduli dimension of that deformation was of  18 \cite{kl1}, not 
the same as the  moduli dimension of algebraic K3 surfaces.
Here, we recovered the same moduli dimension of deformation, 19 by 
algebraically deforming a pair of algebraic K3 surfaces
in a manner similar to
the Connes-Lott construction \cite{cl90}.

Then we considered the extension of this method to the N-point case.
Notice that in the N-point deformation case,
$t_j$ in the 
$t_it_j +t_j t_i= P_{ij}(x_1, x_2)$ 
is not a function on
the $i$-th copy of commutative K3 surface
 $X_i$ for $i \neq j$.
Rather this can be thought of as noncommutative
deformation of $N$ K3 surfaces or  a 
noncommutative deformation of the 
ambient space 
 ${\mbb P}^2 (x_1, x_2) \times {\mbb P}^1 (t_1) \times \cdots \times {\mbb P}^1 (t_N)$
compatible to the
complex structure of each K3 surface.
In the N-point case, we obtained 19N(N-1)/2 as the dimension of deformation moduli. 

When N=3, it is interesting whether we can find an analogue
of the classical hyperk\"{a}hler structure of K3 surface.
  First, we recall the property of
the moduli space of Ricci flat metrics on a K3 surface $S$. 
If a given metric $g$ satisfies $g(Jv,Jw) =g(v,w)$ for any tangent vector $v,w$, then we 
say that the metric $g$ is compatible with the complex structure $J$.
If the two form $\Omega ( \cdot , \cdot ) = g (J \cdot , \cdot ) $ is closed,
then it is called a K\"{a}hler metric and $\Omega$ is called a K\"{a}hler form. 
Any given Ricci-flat metric $g$ induces a
Hodge $*$ operator on $H^2(S,\R) \cong {\R}^{3,19} $ by which $H^2(S,\R)$ 
can be decomposed as a direct sum of two eigenspaces, self dual part (eigenvalue 1)
of dimension 3 and anti-self dual part (eigenvalue $-1$) of dimension 19.

In this setting, for the given Ricci-flat metric $g$, the self dual part $\Lambda^+$
is a 3-dimensional real vector space consisting of vectors whose self intersection is positive.
Different compatible structrues $J$ to $g$ correspond to different unit vectors in $\Lambda^+$,
and they form $S^2$ isomorphic to ${\mbb P}^1$.
Here we choose 3 orthogonal unit vectors $\Omega_1, \Omega_2, \Omega_3$ in $\Lambda^+$
such that corresponding complex structures $J_1, J_2, J_3$ satisfy the relation
$J_i J_j = \epsilon_{ijk} J_k$ for $i,j,k=1,2,3$.
This is called a hyperk\"{a}hler structure on $S$.
We wonder whether we can see  the 3-point deformaiton case
as the deformation of this hyperk\"{a}hler structure on $S$
by relating $t_i$'s with $J_i$'s.

Fr\"{o}hlich {\it et al} \cite{frol1,frol2} defined a spectral triple for 
this hyperk\"{a}hler case introducing the operators 
$\partial, \overline{\partial}, T^i, \overline{T}^i, ~~ i=1,2,3$ acting on the
differential forms.
Here, $\partial = \frac{1}{2}(D -i\overline{D}),$
where $D$ is the Dirac operator and $T^i, ~ i=1,2,3$ are
operators coming from the hyperk\"{a}hler structure.
Then they extended this definition to the noncommutative case.
We also wonder whether we can relate our $t_i$ with their $T_i$.  

 Finally, we wonder whether we can find a sort of Clifford structures
on  ${\mbb P}^2 (x_1, x_2)$  with
the fiber
$ {\mbb P}^1 (t_1) \times \cdots \times {\mbb P}^1 (t_N)$.
This may be considered by regarding 
$ t_i t_j + t_j t_i = P_{ij}(x_1, x_2), ~ i, j = 1, \cdots, N, ~ i \neq j $ and 
$ t_i^2  =  h_i(x_1, x_2),  ~ i=1, \cdots, N, $ not
as constraints giving noncommutative deformation and complex structures
for K3 surfaces but 
as the components of a symmetric matrix giving the metric  on the fiber.

\pagebreak

\vspace{5mm}
\noindent
{\Large \bf Acknowledgments}

\vspace{5mm}
This work was supported by
KOSEF Interdisciplinary Research Grant No. R01-2000-00022.



\newcommand{\MPL}{Mod.\ Phys.\ Lett.}
\newcommand{\NP}{Nucl.\ Phys.}
\newcommand{\PL}{Phys.\ Lett.}
\newcommand{\PR}{Phys.\ Rev.}
\newcommand{\PRL}{Phys.\ Rev.\ Lett.}
\newcommand{\CMP}{Commun.\ Math.\ Phys.}
\newcommand{\JMP}{J.\ Math.\ Phys.}
\newcommand{\JHEP}{JHEP}
\newcommand{\ib}{{\it ibid.}}



\begin{thebibliography}{88}

\bibitem{cds} A. Connes, M.R. Douglas, and A. Schwarz,
{\JHEP} {9802}, 003 (1998), hep-th/9711162.

\bibitem{conn} A. Connes, {\it Noncommutative geometry} (Academic Press, New York, 1994).

\bibitem{cr} A. Connes and M. Rieffel,
{Contemp. Math.} 62,  237 (1987).

\bibitem{rief} M. Rieffel,
{Can. J. Math.} Vol. XL, 257 (1988).

\bibitem{nct} D. Brace, B. Morariu, and B. Zumino,
{\NP} {B 545}, 192 (1999),  hep-th/9810099;
  P.-M. Ho, Y.-Y. Wu, and Y.-S. Wu,
{\PR} {D 58}, 026006 (1998), hep-th/9712201.

\bibitem{hv} C. Hofman and E. Verlinde,
{\NP} {B 547}, 157 (1999), hep-th/9810219.

\bibitem{t4ours}
 E. Kim, H. Kim, N. Kim, B.-H. Lee, C.-Y. Lee, and H. S. Yang,
{\PR} {D 62}, 046001 (2000),  hep-th/9912272.

\bibitem{sw} See, for instance, N. Seiberg and E. Witten,
{\JHEP} {9909},  032 (1999), hep-th/9908142
and references therein for the development in this direction.

\bibitem{bl1} D. Berenstein and R. G. Leigh, Phys. Lett. B 499,  207 (2001),
hep-th/0009209.

\bibitem{kl1} H. Kim and C.-Y. Lee, ``Noncommutative K3 surfaces", hep-th/0105265.

\bibitem{bs} A. Belhaj and E. H. Saidi, ``On noncommutative Calabi-Yau hypersurfaces", hep-th/0108143.

\bibitem{ks1} A. Konechny and A. Schwarz,
{\NP} {B 591},  667 (2000), hep-th/9912185.

\bibitem{ks2} A. Konechny and A. Schwarz,
{\JHEP} {0009},  005 (2000), hep-th/0005174.

\bibitem{kkl} E. Kim, H. Kim, and C.-Y. Lee,
 J. Math. Phys. 42,  2677 (2001), hep-th/0005205.

\bibitem{schwarz} A. Schwarz, Lett. Math. Phys. 58,  81 (2001).    

\bibitem{manin} Y. Manin,  ``Theta functions, quantum tori and Heisenberg groups",
math.AG/0011197.


\bibitem{bjl} D. Berenstein, V. Jejjala, and R. Leigh, Nucl. Phys. B 589,  196 (2000), hep-th/0005087;
 Phys. Lett. B 493,  162 (2000), hep-th/0006168.

\bibitem{vafa} C. Vafa, Nucl. Phys. B 273,  592 (1986).

\bibitem{vfwt} C. Vafa and E. Witten, J. Geom. Phys. 15,  189 (1995), hep-th/9409188.

\bibitem{doug} M. R. Douglas, ``D-branes and discrete torsion", hep-th/9807235.

\bibitem{dgfl} M. R. Douglas and B. Fiol,
``D-branes and discrete torsion II", hep-th/9903031.

\bibitem{jgomi} J. Gomis, JHEP 0005,  006 (2000), hep-th/0001200.

%
%
%

\bibitem{aspinwall} P. S. Aspinwall, ``K3 surfaces and string duality", TASI-96 lecture notes,
  hep-th/9611137.

\bibitem{mukai} S. Mukai,  ``On the moduli space of bundles on K3 surfaces, I'' in
{\it Vector Bundles on Algebraic Varieties,} edited by
M. Atiah {\it et al.} (Oxford Univ. Press, Oxford, 1985).

\bibitem{cl90} A. Connes and J. Lott, Nucl. Phys. B (Proc. Suppl.) 18B,  29 (1990).

\bibitem{frol1} J. Fr\"{o}hlich, O. Grandjean, and A. Recknagel, Commun. Math. Phys. 193,  527 (1998). 

\bibitem{frol2} J. Fr\"{o}hlich, O. Grandjean, and A. Recknagel,  Commun. Math. Phys. 203,  119 (1999).


\end{thebibliography}
\end{document}